\documentclass[a4paper]{article}

\usepackage{INTERSPEECH2022}
\usepackage{enumitem}
\usepackage{multirow}
\usepackage{makecell}
\usepackage{url}
\usepackage{color, colortbl}
\definecolor{Gray}{gray}{0.9}
\newcolumntype{g}{>{\columncolor{Gray}}c}

\title{The CHiME-7 UDASE task: Unsupervised domain adaptation for conversational speech enhancement}
\name{\mbox{Simon Leglaive$^1$}, \mbox{L\'eonie Borne$^2$}, \mbox{Efthymios Tzinis$^{3*}$}\thanks{$^*$Now with Google.}, \mbox{Mostafa Sadeghi$^4$}, \mbox{Matthieu Fraticelli$^5$}, \mbox{Scott Wisdom$^6$}, \mbox{Manuel Pariente$^2$}, \mbox{Daniel Pressnitzer$^5$}, \mbox{John R. Hershey$^6$}}
\address{
  $^1$CentraleSup\'elec, IETR UMR CNRS 6164, France \qquad $^2$Pulse Audition, France \qquad$^3$University of Illinois at Urbana-Champaign, USA \qquad $^4$Inria, France \qquad $^5$\'Ecole Normale Sup\'erieure, PSL University, CNRS, France \qquad$^6$Google, USA \qquad }
\email{}

\begin{document}

\maketitle
\begin{abstract}
Supervised speech enhancement models are trained using artificially generated mixtures of clean speech and noise signals, which may not match real-world recording conditions at test time. This mismatch can lead to poor performance if the test domain significantly differs from the synthetic training domain. This paper introduces the unsupervised domain adaptation for conversational speech enhancement (UDASE) task of the 7th CHiME challenge. This task aims to leverage real-world noisy speech recordings from the target domain for unsupervised domain adaptation of speech enhancement models. The target domain corresponds to the multi-speaker reverberant conversational speech recordings of the CHiME-5 dataset, for which the ground-truth clean speech reference is unavailable. Given a CHiME-5 recording, the task is to estimate the clean, potentially multi-speaker, reverberant speech, removing the additive background noise. We discuss the motivation for the CHiME-7 UDASE task and describe the data, the task, and the baseline system.
\end{abstract}

\section{Introduction}

Modern speech technologies enable us to connect with each other much beyond standard telephony, for instance through video sharing on social media, remote conferencing, or assistive hearing. For these technologies to be truly effective, they must rely on dependable speech processing algorithms that can work optimally in diverse and uncontrolled acoustic environments. Unfortunately, recordings of speech in real-life situations are inevitably contaminated by unwanted background noise, necessitating the use of speech enhancement algorithms to improve the quality and intelligibility of speech.

The speech enhancement task is to estimate a clean speech signal from a noisy recording \cite{loizou2013speech}. In recent years, there has been great progress in speech enhancement thanks to the use of deep learning algorithms. Most speech enhancement models today rely on deep neural networks that are trained in a supervised manner \cite{wang2017supervised}. Supervised speech enhancement requires a training dataset consisting of noisy speech signals and their corresponding clean reference signals. We say that the noisy signals are labeled with the clean speech signals, which are used as the training targets for the speech enhancement model. Given the impossibility of acquiring such labeled data in real conditions, datasets are generated artificially by creating synthetic mixtures of isolated speech and noise signals. However, it is difficult if not impossible to generate a synthetic training dataset that matches arbitrary acoustic conditions at test time, in terms of noise type and level, recording equipment, speaker-to-microphone distance, reverberation, etc. Artificially generated training data are thus inevitably mismatched with real-world noisy speech recordings, which can result in poor speech enhancement performance in case of severe mismatch. Supervised speech enhancement models are therefore domain-specific; if the test domain deviates from the synthetic training domain, it will be necessary to rebuild a training dataset and retrain the model. These limitations of supervised deep learning methods for speech enhancement contrast with the impressive adaptability of the human auditory system when it comes to perceiving speech in unknown adversary acoustic conditions \cite{bent2009perceptual, brandewie2010prior,cooke2022time}. They also contrast with early approaches to supervised speech enhancement, e.g., based on hidden Markov models \cite{sameti1998hmm} or nonnegative matrix factorization \cite{mohammadiha2013supervised}, in which adaptation to unseen acoustic conditions was part of the proposed methodologies.

Previous data challenges for single-channel speech enhancement have focused on such supervised setups, where labeled training data are provided that match the evaluation domain. Such challenges include the series of deep noise suppression (DNS) Challenges \cite{reddy2020interspeech, reddy2021icassp, reddy2021interspeech, dubey2022icassp}, which also developed the non-intrusive DNSMOS \cite{reddy2022dnsmos} model for automatic quality assessment. The DNS Challenge training sets have consisted of a large amount of multi-condition data intended to cover diverse conditions. Such approaches can be effective, so long as test-time conditions are covered by the training data. In contrast, the proposed challenge is intended to study a different situation where we are targeting single-channel speech enhancement in a specific domain for which no well-matched labeled data are available for training.
% TODO: maybe talk about Clarity challenge? But it is multimic...
% TODO: Maybe talk about previous CHiME challenges? The goal is generally different for those: optimizing ASR performance. Also, multimic
% TODO: ConferencingSpeech 2021 (https://tea-lab.qq.com/conferencingspeech-2021), but this is also multimic?

Recording unlabeled noisy speech signals in the target domain is much easier than engineering synthetic clean speech and noise mixtures that match this domain. However, leveraging such unlabeled data to develop a speech enhancement model is a challenging problem, which is the focus of the unsupervised domain adaptation for conversational speech enhancement (UDASE) task of the CHiME-7 challenge. The problem we propose to address in this task consists of using unlabeled data in the target domain to adapt supervised speech enhancement models trained on synthetic labeled data in a mismatched source domain. This corresponds to an unsupervised domain adaptation task, but the general problem of improving the generalization capability of speech enhancement models to real-world conditions for which labeled data are not available can also be addressed using fully unsupervised or semi-supervised learning algorithms.

In the CHiME-7 UDASE task, the target domain corresponds to the real conversational speech recordings of the CHiME-5 dataset \cite{barker2018fifth}. These recordings were made during dinner parties, so they include multiple speakers having a natural conversation in noisy and reverberant environments.
For supervised learning in a mismatched source domain, we rely on the artificially generated LibriMix dataset \cite{cosentino2020librimix}, which is derived from LibriSpeech clean utterances \cite{panayotov2015librispeech} and WHAM! noises \cite{wichern2019wham}. The CHiME-7 UDASE task consists of denoising CHiME-5 conversational speech recordings using the LibriMix out-of-domain (OOD) labeled data and the CHiME-5 in-domain unlabeled data. Given a mixture of one or more reverberant speakers and additive noise, the goal is to predict the clean audio signal of the reverberant speaker(s), removing the additive noise. This task is motivated by the assistive listening use case, in which a speech enhancement algorithm can help any individual to better engage in a conversation, by improving the overall multi-speaker speech intelligibility and quality within the ambient noise. For development and evaluation only, we release the reverberant LibriCHiME-5 dataset, which consists of synthetic mixtures generated to be close to the target domain. Systems submitted to the CHiME-7 UDASE task will be first evaluated using objective performance metrics, then the best-performing systems will be evaluated through a subjective listening test following the ITU-T recommendation P.835~\cite{recommendation2003subjective}.

The tools and resources provided to the participants of the CHiME-7 UDASE task are available in the repositories of our GitHub organization\footnote{\url{https://github.com/UDASE-CHiME2023}} and at the CHiME challenge website\footnote{\url{https://www.chimechallenge.org/}}.

The paper is structured as follows. The datasets and the task to be solved are described in Sections~\ref{sec:data} and \ref{sec:task}. The baseline is presented in Section~\ref{sec:baseline}. We conclude in Section~\ref{sec:conclusion}.

\section{Data}
\label{sec:data}

The CHiME-7 UDASE task builds upon the following datasets that will be presented in the next subsections: the CHiME-5 in-domain unlabeled data for training, development and evaluation \cite{barker2018fifth}; the LibriMix OOD labeled data for training and development \cite{cosentino2020librimix}; the reverberant LibriCHiME-5 close-to-in-domain labeled data for development and evaluation.

\subsection{CHiME-5 in-domain unlabeled data}
\label{subsec:chime-5}

The CHiME-5 data consists of twenty 4-people dinner parties or sessions, of between two and three hours, each recorded in a different home with three recording locations per home (kitchen, dining room, living room) \cite{barker2018fifth}. The audio recordings include natural conversations between multiple speakers in reverberant and noisy environments, and these are fully transcribed. Using the CHiME-5 transcription files, we estimated that 22\% of the audio recordings contain only noise, 51\% contain one single active speaker, and 20\%, 5\% and 2\% contain two, three and four overlapping speakers, respectively (numbers obtained without considering any constraint on the overlap duration). 

For the CHiME-7 UDASE task, we only use the right channel of the binaural recordings because the left channel is less reliable. We also discard portions of the data where the participant wearing the microphone speaks, to simulate an assistive listening situation where all voices but the listener's should be enhanced.
The extracted audio segments therefore contain up to three simultaneously-active reverberant speakers and background noise. The noisy speech signals are not labeled with the clean speech reference signals. The main objective of the UDASE task is to develop new approaches that can leverage this in-domain unlabeled dataset for speech enhancement.

The training set consists of raw single-channel audio segments extracted from the binaural recordings. Developing and evaluating a speech enhancement model requires computing objective performance metrics. This is a difficult problem as the CHiME-5 dataset contains noisy multi-speaker speech recordings that are not labeled with clean speech reference signals. For development and evaluation, we therefore used the transcription of the CHiME-5 recordings to extract short audio segments of duration at least 3 seconds labeled with the maximum number of simultaneously-active speakers (0, 1, 2 or 3),\footnote{This only corresponds to a maximum value, i.e., through the duration of a segment the number of simultaneously-active speakers can vary between 0 and the maximum value. Moreover, a segment might contain more speakers than the labeled maximum number of simultaneously active speakers. For instance, a segment labeled as single-speaker might contain two active speakers who do not speak simultaneously.} which will allow us to compute objective performance metrics. This procedure simulates the reasonable scenario where we can afford to manually annotate a small amount of data with speaker count labels for development and evaluation, but this procedure cannot be easily done for a large training set. The audio segment extracted for the development and evaluation sets is done as follows: (i) we extract all segments where no speaker is active (i.e., noise-only segments); (ii) we extract all segments that were not extracted previously and without overlapping speakers (i.e., single-speaker segments); (iii) we extract all segments that were not extracted previously and with at most two overlapping speakers; (iv) we extract all segments that were not extracted previously and with at most three overlapping speakers. As a post-processing, Brouhaha \cite{lavechin2022brouhaha} was used on the 0- and 1-speaker segments to verify the absence and presence of speech, respectively. Misclassified segments were reviewed and removed when appropriate. Noise-only segments are used to create the reverberant LibriCHiME-5 dataset (see Section~\ref{subsec:librichime-5}), allowing for computing objective performance metrics such as the scale-invariant signal-to-distortion ratio (SI-SDR) \cite{leroux2019sdr} on close-to-in-domain data. Single-speaker segments will be used to compute DNSMOS P.835 (simply referred to as DNSMOS hereinafter) metrics \cite{reddy2022dnsmos}. There is no official guideline on how to use the 2- and 3-speaker segments.

We also provide an evaluation subset that will be used for the listening test. It consists of audio samples that were extracted by looking for segments of 4 to 5 seconds with at least 3 seconds of speech and 0.25 second without speech at the beginning and at the end. Additional constraints were taken into account to ensure a balanced subset in terms of speaker gender, recording location, and session.

\subsection{LibriMix out-of-domain labeled data}
\label{subsec:librimix}

For supervised learning on OOD data, we chose the LibriMix dataset \cite{cosentino2020librimix} because it is a standard open-source dataset in the community. LibriMix was originally developed for speech separation in noisy environments, it is derived from LibriSpeech clean utterances \cite{panayotov2015librispeech} and WHAM! noises \cite{wichern2019wham}. The Libri2Mix and Libri3Mix versions of the dataset contain noisy speech mixtures with 2 and 3 overlapping speakers, respectively. A single-speaker version of LibriMix (Libri1Mix) can be obtained by simply discarding one of the two speakers in Libri2Mix mixtures. For a complete description of LibriMix, the interested reader is referred to \cite{cosentino2020librimix}.

\subsection{Reverberant LibriCHiME-5 close-to-in-domain labeled data}
\label{subsec:librichime-5}

In real-world conditions, in particular for the CHiME-5 recordings, it is impossible to have access to the ground-truth clean speech reference signals associated with the noisy recordings due to cross-talk between microphones. Yet, when developing and evaluating a speech enhancement algorithm, it is necessary to compute objective performance metrics. For this purpose, we created the reverberant LibriCHiME-5 dataset for development and evaluation only.
This dataset consists of synthetic mixtures of reverberant speech and noise, with up to three simultaneously active speakers, labeled with the clean reference speech signals. Noise signals were extracted from the CHiME-5 recordings using the ground-truth transcriptions, and clean speech utterances were taken from the LibriSpeech dataset \cite{panayotov2015librispeech} and were convolved with room impulse responses (RIRs) from the VoiceHome corpus \cite{bertin2016french}. These RIRs were recorded in 12 different rooms of 3 real homes, with 4 rooms per home: living room (room 1), kitchen (room 2), bedroom (room 3), and bathroom (room 4). Bathrooms were excluded for the reverberant LibriCHiME-5 dataset. In each room, RIRs were recorded for 2 different positions and geometries of an 8-channel microphone array and 7 to 9 different positions of the loudspeaker. 

For each mixture in the reverberant LibriCHiME-5 dataset, we randomly choose the maximum number $n \in \{1,2,3\}$ of simultaneously-active speakers in the mixture, with $p(n=i) = 0.60, 0.35, 0.05$ for $i=1,2,3$, respectively, which is consistent with the distribution of the segmented CHiME-5 dataset. In the VoiceHome corpus, we randomly and successively sample a home, a room, an array position/geometry, $n$ speaker positions without replacement, and a channel of the microphone array, which gives the RIRs for the current mixture. LibriSpeech utterances are convolved with the selected RIRs to obtain the reverberant speech utterances. These are mixed following speech activity patterns extracted from the CHiME-5 transcription files to simulate a natural conversation between multiple speakers. Multi-speaker reverberant speech and noise mixtures are created such that the per-speaker signal-to-noise ratio (SNR) is distributed as a Gaussian with a mean of 5~dB and a standard deviation of 7~dB, to match the SNR distribution of the CHiME-5 dataset as estimated by Brouhaha \cite{lavechin2022brouhaha}. This is achieved by first sampling a global per-mixture SNR $x \sim \mathcal{N}(5, \sigma_1^2)$ and then sampling a local per-speaker SNR $y \sim \mathcal{N}(x, \sigma_2^2)$, with $\sigma_1 = 6.7082$ and $\sigma_2 = 2$ ($\sqrt{\sigma_1^2 + \sigma_2^2} \approx 7$~dB). The value of $\sigma_2$ is chosen such that the loudness difference between multiple speakers is moderate, this is again to simulate a conversation.

Despite the effort to generate a synthetic dataset that matches the distribution of the target domain as much as possible, there still exists a mismatch between the reverberant LibriCHiME-5 dataset and the CHiME-5 dataset, e.g., read speech for the latter and spontaneous speech for the former; the reverberation times might also differ. It is indeed impossible to create synthetic labeled data that perfectly match real-world unlabeled recordings, hence the CHiME-7 UDASE task. Nevertheless, as already mentioned, it is required for development and evaluation to be able to compute objective performance metrics, complementary to listening tests. DNSMOS \cite{reddy2022dnsmos} provides a way to evaluate the performance on single-speaker segments of the CHiME-5 data without having access to the clean speech reference signals, but this is not sufficient as a non-negligible proportion of the CHiME-5 data contains simultaneously-speaking people. We believe it is reasonable to expect systems that successfully managed to leverage the unlabeled CHiME-5 data to have better results on the reverberant LibriCHiME-5 dataset than fully supervised systems only trained on the labeled LibriMix dataset. Indeed, in the reverberant LibriCHiME-5, the speech utterances were convolved with real RIRs measured in domestic environments, the noise signals were extracted from the CHiME-5 recordings, the per-speaker SNR was chosen to approximately match that of the CHiME-5 data, and the speech utterances were mixed to simulate a conversation using the CHiME-5 transcription. We can thus hope that the performance computed on the reverberant LibriCHiME-5 dataset corresponds to an imperfect estimate of the performance on the CHiME-5 dataset. 

\begin{table}[t]
\centering
\resizebox{1.0\linewidth}{!}{ 
\begin{tabular}{ccccc}
\cmidrule[.8pt]{1-5}
& & \multicolumn{2}{c}{\makecell{Sample length (s)}}          & Total duration \\
Subset & \# samples & Mean & STD &     (HH:MM:SS)                                       \\
\cmidrule(lr){1-1} \cmidrule(lr){2-2} \cmidrule(lr){3-3} \cmidrule(lr){4-4} \cmidrule(lr){5-5}
\texttt{train}           & 27 517                                                  & 10.91                    & 14.10                   & 83:22:29                                                       \\ \cmidrule(lr){1-5}
\texttt{dev/0}           & 912                                                    & 6.50                     & 4.10                    & 1:38:49                                                        \\
\texttt{dev/1}           & 5 719                                                   & 5.89                     & 3.49                    & 9:21:53                                                        \\
\texttt{dev/2}           & 3 835                                                   & 5.23                     & 2.43                    & 5:34:33                                                        \\
\texttt{dev/3}           & 667                                                    & 4.61                     & 1.84                    & 0:51:14                                                        \\ \cmidrule(lr){1-5}
\texttt{eval/0}          & 977                                                    & 5.73                     & 3.35                    & 1:33:19                                                        \\
\texttt{eval/1}          & 3 013                                                   & 5.54                     & 2.94                    & 4:35:05                                                        \\
\texttt{eval/2}          & 1 552                                                   & 4.88                     & 2.04                    & 2:06:07                                                        \\
\texttt{eval/3}          & 233                                                    & 4.21                     & 1.17                    & 0:16:21                                                        \\
\texttt{eval/LT} 		& 241                                                    & 4.72                     & 0.34                    & 0:18:58   \\ \cmidrule[.8pt]{1-5}
\end{tabular}
}
\vspace{.1cm}
\caption{Segmented CHiME-5 dataset. Dev and eval subsets are labeled with the maximum number of simultaneously active speakers (0, 1, 2, 3). \texttt{eval/LT} corresponds to the evaluation subset for the listening test.}
\label{tab:chime-5}
\vspace{-.5cm}
\end{table}
\begin{table}[t]
\centering
\resizebox{1.0\linewidth}{!}{ 
\begin{tabular}{ccccc}
\cmidrule[.8pt]{1-5}
& & \multicolumn{2}{c}{\makecell{Sample length (s)}}          & Total duration \\
Subset & \# samples & Mean & STD &     (HH:MM:SS)                                        \\
\cmidrule(lr){1-1} \cmidrule(lr){2-2} \cmidrule(lr){3-3} \cmidrule(lr){4-4} \cmidrule(lr){5-5}
\texttt{dev/1}  & 1 187              & 7.14               & 4.67               & 2:21:09        \\
\texttt{dev/2}  & 565               & 5.37               & 2.24               & 0:50:31        \\
\texttt{dev/3}  & 65                & 4.81               & 1.66               & 0:05:12        \\ \cmidrule(lr){1-5}
\texttt{eval/1} & 1 394              & 6.25               & 3.75               & 2:25:17        \\
\texttt{eval/2} & 494               & 4.44               & 1.34               & 0:36:35        \\
\texttt{eval/3} & 64                & 4.21               & 1.07               & 0:04:29                                                       \\ \cmidrule[.8pt]{1-5}
\end{tabular}
}
\vspace{.1cm}
\caption{Reverberant LibriCHiME-5 dataset. The subsets are labeled with the maximum number of simultaneously active speakers (0, 1, 2, 3).
}
\label{tab:librichime-5}
\vspace{-.5cm}
\end{table}

\section{Task}
\label{sec:task}

\subsection{Training, development, and evaluation sets}
\label{subsec:data_splits}

In the original CHiME-5 dataset \cite{barker2018fifth}, the 20 sessions (or dinner parties) were divided into disjoint training (train), development (dev), and evaluation (eval) sets \cite{barker2018fifth}. For the CHiME-7 UDASE task, we move sessions S07 and S17 from the train set to the dev set to obtain a sufficient amount of noise-only segments for the generation of the reverberant LibriCHiME-5 dataset. There is no overlap between speakers in each set. The segmented CHiME-5 dataset for the CHiME-7 UDASE task is summarized in Table~\ref{tab:chime-5}.

The dev (resp.\ eval) set of the reverberant LibriCHiME-5 dataset is created from the `dev-clean' (resp.\ `test-clean') subset of LibriSpeech, noise-only segments from the dev (resp.\ eval) set of CHiME-5, and a subset of VoiceHome RIRs. RIRs from home 2 (rooms 1, 2, 3) and home 3 (rooms 1, 3) are used for the dev set, and RIRs from home 3 (room 2) and home 4 (rooms 1, 2, 3) are used for the eval set. The reverberant LibriCHiME-5 dataset is summarized in Table~\ref{tab:librichime-5}.

The original split of the LibriMix dataset \cite{cosentino2020librimix} into train, dev, and eval (test) subsets is kept for the CHiME-7 UDASE task.

\subsection{Rules}

The train and dev sets of the LibriSpeech and WHAM! datasets (from which LibriMix is generated) can be used individually (e.g., to train isolated speech and noise models) or they can be used to create synthetic mixtures similar to the original LibriMix dataset. Specifically, participants are allowed to create synthetic mixtures using noise-only segments that would be extracted from the binaural recordings of the CHiME-5 training set, only if this extraction does not rely on the CHiME-5 ground-truth transcription.
Participants are also allowed to use RIRs to create reverberant utterances from LibriSpeech, as long as the RIRs are synthetic. Using any other datasets of clean speech signals, noise signals, or measured RIRs is not allowed.
Finally, the Kinect recordings of the CHiME-5 dataset cannot be used.
Although a synthetic labeled dataset better matching with the real CHiME-5 data could be created with more engineering effort and knowledge about the target domain, the goal of the CHiME-7 UDASE task is to simulate more realistic conditions where knowledge about in-domain data is scarce. The motivation for the above rules is to encourage participants to use a relatively identical synthetic dataset and show that models trained with OOD labeled data can be adapted using unsupervised, self- or semi-supervised learning from in-domain unlabeled data.

All speech enhancement system parameters should be tuned on the training set or development set of the LibriMix, CHiME-5 and reverberant LibriCHiME-5 datasets as described in Section~\ref{subsec:data_splits}, or variations that comply with the above rules.
During evaluation or inference, the submitted systems must use as input only noisy speech waveforms and process them independently of one another. 
Participants can use external pre-trained and frozen models for voice activity detection, diarization, speaker counting, or signal-to-noise ratio estimation.

\subsection{Evaluation}
\label{subsec:evaluation}

The submitted systems will follow a two-step evaluation process. They will first be evaluated in terms of SI-SDR \cite{leroux2019sdr} on the complete eval set of the reverberant LibriCHiME-5 dataset and in terms of DNSMOS scores on the \texttt{eval/1} subset of the CHiME-5 dataset. DNSMOS is a non-intrusive objective metric that provides performance scores for the speech signal quality (SIG), the background intrusiveness (BAK), and the overall quality (OVRL) \cite{reddy2022dnsmos}. The four best-performing systems in terms of SI-SDR or OVRL score will then be evaluated by a listening test using audio samples from the \texttt{eval/LT} subset of the CHiME-5 dataset. In case a team submits multiple entries, only the one that obtains the best performance during the first evaluation stage will be qualified for the listening test. 

The evaluation data will be released two weeks before the submission deadline. Participants are asked to evaluate their system using the provided evaluation scripts, and they are asked to return the performance scores for each audio file of the \texttt{eval/1} subset of the CHiME-5 dataset and of the \texttt{eval/\{1,2,3\}} subsets of the reverberant LibriCHiME-5 dataset. They are also asked to submit the output signals of their system to allow their scores to be verified, and a technical report describing their system. Participants are asked to normalize the output signals at a loudness of -30 LUFS (Loudness Unit Full Scale) before computing the DNSMOS performance scores, using the Python package \texttt{pyloudnorm} \cite{steinmetz2021pyloudnorm}. The motivation for this normalization is that DNSMOS scores (especially the SIG and BAK scores) are very sensitive to a change in the input signal loudness. This sensitivity would make it difficult to compare different systems without a common normalization procedure. The same normalization is considered for the listening test material.

Optionally, participants are also invited to submit SI-SDR scores for the LibriMix dataset (`max' version), using the \texttt{test/mix\_single} and \texttt{test/mix\_both} subsets of Libri2Mix (containing 3000 single-speaker and 2-speaker examples, respectively) and the \texttt{test/mix\_both} subset of Libri3Mix (containing 3000 3-speaker examples). SI-SDR results on LibriMix will not be used to rank systems because it would not be consistent with the purpose of the CHiME-7 UDASE task. They will only be used to compare the performance on the (close to) in-domain and OOD datasets.

The listening test will follow the ITU-T Recommendation P.835 \cite{recommendation2003subjective}. It will be conducted in person in a listening booth at the University of Sheffield. Participants will listen over headphones to short speech samples (4–5 seconds). Each trial will consist of three presentations of the same sample, to collect three different subjective reports. In the different presentations participants will be instructed to either focus on the speech signal and rate how natural it sounds, focus on the background noise and rate how noticeable or intrusive this background is, or attend to both the speech and the background noise and rate the overall quality of the sample, quality being defined in the perspective of everyday speech communication. The order of presentations will be counterbalanced across participants. The ratings will be reported on 5-point Likert scales and mean opinion scores (MOS) will be computed. 

We target a total number of 32 subjects, separated into 4 panels of 8 listeners. Each panel will be associated with a distinct set of 32 audio samples taken from the \texttt{eval/LT} subset of the CHiME-5 dataset, resulting in a total of 32 $\times$ 4 = 128 audio samples for the entire listening test. For each audio sample we will have 5 different experimental conditions (4 systems and the noisy input condition). Each listener will evaluate all experimental conditions for each audio sample associated with his/her panel, according to the three rating scales. For each pair of audio sample and experimental condition, a MOS will be computed out of 8 votes, leading to an overall 8 $\times$ 128 = 1024 votes for each experimental condition.
The final ranking of the systems will be based on statistical analysis of the MOS results. 

\section{Baseline}
\label{sec:baseline}

\subsection{Baseline system}

The CHiME-7 UDASE baseline system is based on RemixIT \cite{tzinis2022continual,tzinis2022remixit}, a self-supervised learning approach for unsupervised domain adaptation of a speech enhancement model pre-trained (in a supervised or a self-supervised manner) on OOD noisy speech data. In semi-supervised RemixIT, a teacher model is trained on OOD noisy speech signals alongside the corresponding clean speech and noise reference waveforms. In the second step, RemixIT performs inference on a batch of noisy in-domain speech recordings to obtain pseudo-labels that are used to train a student model for speech enhancement in the target domain without the need for in-domain reference signals. 

\subsubsection{OOD Sudo rm -rf teacher model} 

The teacher is based on a Sudo rm -rf \cite{tzinis2020sudo,tzinis2022compute} sound separation model, which is an end-to-end framework with three main blocks: (i) an encoder network processing the raw waveform of the input audio mixture; (ii) a separator network that operates on the encoder output to provide separation masks; (iii) a decoder network to estimate the audio source signals from the encoder output and the estimated masks. The encoder and decoder architectures consist of a one-dimensional convolution and transpose convolution, respectively, with 512 filters of 41 taps and a hop size of 20 samples. The backbone structure of the separator network consists of 8 U-Conv blocks where each block extracts and aggregates information from multiple resolutions.

The Sudo rm -rf teacher model is trained fully supervised on the LibriMix OOD labeled dataset. We use the 3-speaker version of the dataset (Libri3Mix) and discard one (two) speakers in the original Libri3Mix mixtures to obtain 2- (single-) speaker data. The proportion of single-speaker, 2-speaker, and 3-speaker mixtures are 0.50, 0.25, and 0.25, respectively. The model is trained by minimizing the negative SI-SDR loss for both the speech and the noise components with equal weights.

\subsubsection{Self-supervised RemixIT student model} 

The student model follows the same architecture as the teacher model, and it is initialized using the model parameters resulting from the supervised training on LibriMix. The RemixIT learning framework consists of feeding noisy speech signals from the training set of the unlabeled CHiME-5 dataset in the frozen teacher model to get estimates of the isolated speech and noise, which will serve as pseudo-labels. In a second step, new bootstrapped mixtures are synthesized by permuting the aforementioned noise estimates and remixing them with the speech estimates. These bootstrapped mixtures and the corresponding pseudo-labels are finally used to train the student model, again by minimizing the SI-SDR loss between the student's speech and noise estimates and the corresponding teachers' pseudo-targets. The teacher model is also continuously updated using the parameters of the student model, following an exponential moving average update. The final student model is chosen as the one obtaining the highest mean overall MOS (OVRL) as computed by DNSMOS on the single-speaker subset of the CHiME-5 dev set (\texttt{dev/1} subset).

We provide two versions of the student model. The first one is trained on the raw audio segments of the CHiME-5 \texttt{train} set, which may result in sub-optimal performance because these audio segments do not always contain speech. Remixing possibly perturbed noise waveforms with almost zero teacher's speech estimates will not always lead to valid new bootstrapped mixtures for training the student model. Therefore, we provide a second student model that was trained on audio segments of the CHiME-5 \texttt{train} set which have been automatically labeled as containing speech by Brouhaha's voice activity detector (VAD)~\cite{lavechin2022brouhaha}.

\subsection{Results and discussion}

\begin{table}[t]
\resizebox{1.0\linewidth}{!}{ 
\begin{tabular}{cccccc}
\cmidrule[.8pt]{1-6}
Subset & Metric & Input & OOD teacher & RemixIT & RemixIT-VAD\\ \cmidrule(lr){3-6}
& & \multicolumn{4}{c}{LibriMix dataset} \\\cmidrule(lr){3-6}
\texttt{dev} & \multirow{2}{*}{SI-SDR} & 5.2 & \textbf{13.2} & 11.9 & 12.3 \\ 
\texttt{eval} & & 4.9 & \textbf{13.2} & 11.5 & 12.2 \\ \cmidrule(lr){3-6}
& & \multicolumn{4}{c}{CHiME-5 dataset} \\\cmidrule(lr){3-6}
\multirow{3}{*}{\texttt{dev/1}} & SIG & \textbf{3.64} & 3.48 & 3.44 & 3.46 \\
& BAK & 3.04 & 3.79 & \textbf{3.85} & \textbf{3.85} \\
& OVRL & 3.03 & 3.08 & 3.07 & \textbf{3.09} \\ \cmidrule(lr){3-6}
\multirow{3}{*}{\texttt{eval/1}} & SIG & \textbf{3.48} & 3.33 & 3.26 & 3.28 \\
& BAK & 2.92 & 3.59 & \textbf{3.64} & 3.62 \\
& OVRL & 2.84 & \textbf{2.88} & 2.82 & 2.84 \\\cmidrule(lr){3-6}
& & \multicolumn{4}{c}{Reverberant LibriCHiME-5 dataset} \\\cmidrule(lr){3-6}
\texttt{dev} & \multirow{2}{*}{SI-SDR} & 6.6 & 8.3 & 9.5 & \textbf{9.9} \\ 
\texttt{eval} & & 6.6 & 7.8 & 9.4 & \textbf{10.1} \\ \cmidrule[.8pt]{1-6}
\end{tabular}
}
\caption{Results computed from the unprocessed noisy speech signals (Input) and from the output signals of the three baseline systems (OOD teacher, RemixIT, and RemixIT-VAD).}
\label{tab:results}
\vspace{-.5cm}
\end{table}

Results are shown in Table~\ref{tab:results}. In this table, OOD teacher corresponds to the fully supervised Sudo rm -rf teacher model, while RemixIT and RemixIT-VAD correspond to the student models, the latter being trained on the data preprocessed by Brouhaha's VAD. SIG, BAK, and OVRL correspond to the DNSMOS scores (in between 1 and 5, the higher the better) and are averaged over the 1-speaker subset of the CHiME-5 dev or eval sets. For the reverberant LibriCHiME-5 dataset, the SI-SDR scores (in dB, the higher the better) are averaged over the entire dev or eval sets (including all 1-, \mbox{2-,} and 3-speaker mixtures). This is the same for the LibriMix dataset, where the \texttt{eval} row contains the results averaged over the eval subsets indicated in Section~\ref{subsec:evaluation}, and the \texttt{dev} row contains the results averaged over the equivalent subsets of the LibriMix dev set (i.e., \texttt{dev/\{mix\_single, mix\_both\}} for Libri2Mix and \texttt{dev/mix\_both} for Libri3Mix).

Using RemixIT training, we expect the performance of the speech enhancement student model to improve on the target domain (corresponding to the CHiME-5 data) and to deteriorate on the synthetic training domain (corresponding to the LibriMix data), compared to the OOD teacher model. This is globally what we observe in Table~\ref{tab:results}. In terms of SI-SDR on the LibriMix dataset, the self-supervised student models, namely, RemixIT and RemixIT-VAD perform between 0.9 and 1.7~dB worse than the fully-supervised teacher model as they are getting more fine-tuned towards the target domain. However, as will be discussed in the next paragraphs, the RemixIT student models globally outperform the fully-supervised teacher model on the CHiME-5 and reverberant LibriCHiME-5 datasets.

On the CHiME-5 dataset, the student models obtain better BAK scores compared to the OOD teacher on both the dev and eval set (between +0.03 and +0.06 points), indicating a reduction of the background noise intrusiveness in the output signal. Regarding, the SIG metric, all models obtain lower performance than the unprocessed noisy speech signals, which is expected because these are contaminated by noise but not distorted. Unfortunately, as indicated by the SIG scores the distortion introduced by the speech enhancement process is more important for the student models than for the teacher, which is particularly true for the \texttt{eval/1} subset. 
Regarding the OVRL metric on the \texttt{dev/1} set, the performance of all models is close. RemixIT-VAD outperforms RemixIT and obtains a marginal improvement of 0.01 point compared to the OOD teacher, which we nevertheless considered as sufficient for the baseline of a new challenge. However, the OOD teacher outperforms the student models by a margin between 0.04 and 0.06 points in terms of OVRL score on the CHiME-5 \texttt{eval/1} set, which we assume is mainly due to the increased distortion for the student models. 

Finally, it can be seen that RemixIT's unsupervised adaptation of the OOD Sudo rm -rf teacher model on the in-domain CHiME-5 dataset improved the speech enhancement performance on the close-to-in-domain reverberant LibriCHiME-5 dataset. The student models obtain better performance compared to the fully supervised teacher model, and RemixIT-VAD is the best-performing system with an improvement between 1.6 and 2.3~dB in terms of SI-SDR. Consequently, an unsupervised domain adaptation method that leverages the CHiME-5 data could obtain better separation performance on the reverberant LibriCHiME-5 dataset than a fully-supervised model only trained on the OOD LibriMix data.

\section{Conclusion}
\label{sec:conclusion}

In this paper, we presented the CHiME-7 UDASE task, which aims to foster new methods toward more ecologically valid and robust speech enhancement models. Ecological validity describes the extent to which an experimental setting and task correspond to real-life conditions \cite{hung2019complex}. Fully-supervised speech enhancement models trained (and most of the time also evaluated) on synthetic data cannot always capture the distribution of real-world acoustic recordings, which has important implications in terms of generalization capability. Evaluating unsupervised domain adaptation methods for speech enhancement is by definition a challenging task because one cannot have access to the ground-truth clean speech signals in the target domain. Hopefully, the design of the CHiME-7 UDASE task will enable the further development and evaluation of such methods in the future. We will evaluate the systems submitted to the CHiME-7 UDASE task and the results will be announced during the CHiME-2023 workshop.

\section{Acknowledgments}

The authors thank the CHiME Steering Group (Jon Barker, Emmanuel Vincent, Shinji Watanabe, Michael Mandel, Marc Delcroix, Leibny Paola Garcia Perera) for their support in the organization of the CHiME-7 UDASE task and for their suggestion of using the binaural microphone recordings of the CHiME-5 dataset to create the task material.

\bibliographystyle{IEEEtran}

\bibliography{mybib}

\end{document}